\documentclass[iop]{emulateapj}
\usepackage{amssymb,amsmath}
\usepackage{graphicx}
\usepackage{natbib}

\newcommand{\ciza}{CIZA J2242.8+5301 }
\newcommand{\cizans}{CIZA J2242.8+5301} 
\newcommand{\hms}[3]{{#1}$^\mathrm{h}${#2}$^\mathrm{m}${#3}$^\mathrm{s}$}
\newcommand{\dms}[3]{{#1}\arcdeg{#2}\arcmin{#3}\arcsec}

\slugcomment{LLNL-DOC-661846-DRAFT}

\shorttitle{Imaging and Redshift Analysis of \cizans}
\shortauthors{Dawson et al.}

\begin{document}

\title{MC$^2$: Galaxy Imaging and Redshift Analysis of the Merging Cluster \cizans}

\author{William A. Dawson\altaffilmark{1}, 
M. James Jee\altaffilmark{2},
Andra Stroe\altaffilmark{3},
Y. Karen Ng\altaffilmark{2},
Nathan Golovich\altaffilmark{2},
David Wittman\altaffilmark{2},
David Sobral\altaffilmark{3,4,5},
M. Br\"uggen\altaffilmark{6},
H. J. A. R\"ottgering\altaffilmark{3},
R. J. van Weeren\altaffilmark{7}}

\altaffiltext{1}{Lawrence Livermore National Lab, 7000 East Avenue, Livermore, CA 94550, USA}
\altaffiltext{2}{University of California, One Shields Avenue, Davis, CA 95616, USA}
\altaffiltext{3}{Leiden Observatory, Leiden University, PO Box 9513, NL-2300 RA Leiden, the Netherlands}
\altaffiltext{4}{Instituto de Astrof\'{\i}sica e Ci\^{e}ncias do Espa\c{c}o, Universidade de Lisboa, OAL, Tapada da Ajuda, PT1349-018}
\altaffiltext{5}{Center for Astronomy and Astrophysics of the University of Lisbon, Tapada da Ajuda - Edificio Leste - 2$\arcdeg$ Piso, 1349-018 Lisbon, Portugal}
\altaffiltext{6}{Hamburger Sternwarte, Universit\"at Hamburg, Gojenbergsweg 112, 21029 Hamburg, Germany}
\altaffiltext{7}{Harvard-Smithsonian Center for Astrophysics, 60 Garden Street, Cambridge, MA 02138, USA}

\email{dawson29@llnl.gov}

\date{Draft \today}

\label{firstpage}

\begin{abstract}
X-ray and radio observations of \ciza suggest that it is a major cluster merger.
Despite being well studied in the X-ray, and radio, little has been presented on the cluster structure and dynamics inferred from its galaxy population.
We carried out a deep ($i<25$) broad band imaging survey of the system with Subaru SuprimeCam ($g$ \& $i$ bands) and the Canada France Hawaii Telescope ($r$ band) as well as a comprehensive spectroscopic survey of the cluster area (505 redshifts) using Keck DEIMOS.
We use this data to perform a comprehensive galaxy/redshift analysis of the system, which is the first step to a proper understanding the geometry and dynamics of the merger, as well as using the merger to constrain self-interacting dark matter.
We find that the system is dominated by two subclusters of comparable richness with a projected separation of $6.9\arcmin^{+0.7}_{-0.5}$ (1.3$^{+0.13}_{-0.10}$\,Mpc).
We find that the north and south subclusters have similar redshifts of $z\approx0.188$ with a relative line-of-sight velocity difference of $69\pm 190\,\mathrm{km}\,\mathrm{s}^{-1}$.
We also find that north and south subclusters have velocity dispersions of $1160^{+100}_{-90}\,\mathrm{km}\,\mathrm{s}^{-1}$ and $1080^{+100}_{-70}\,\mathrm{km}\,\mathrm{s}^{-1}$, respectively.
These correspond to masses of $16.1^{+4.6}_{-3.3}\times 10^{14}$\,M$_\sun$ and  $13.0^{+4.0}_{-2.5}\times 10^{14}$\,M$_\sun$, respectively.
While velocity dispersion measurements of merging clusters can be biased we believe the bias in this system to be minor due to the large projected separation and nearly plane-of-sky merger configuration.
\ciza is a relatively clean dissociative cluster merger with near 1:1 mass ratio, which makes it an ideal merger for studying merger associated physical phenomena.
\end{abstract}





\section{Introduction}\label{sec:Intro}
Under the hierarchical structure formation paradigm all clusters are formed from merging substructures.
When the mergers involve two approximately equal mass subclusters a \emph{dissociative} merger can occur where the baryonic plasma of each subcluster collides, forms shocks, is slowed relative to the effectively collisionless galaxies, and becomes dissociated for a time post-merger (some examples include: the Bullet Cluster, \citealt{Clowe:2004eq}; the Musket Ball Cluster, \citealt{Dawson:2012dl}; and Pandora's Cluster, \citealt{merten:2011gu}).
These plasma shocks can lead to sharp X-ray bow shock features \citep{Markevitch:2002iz, Markevitch:2005bg}, and coupled with  the intra-cluster magnetic fields can lead to radio relics, which are diffuse synchrotron sources typically at the periphery of cluster mergers \citep[see][for a review]{Feretti:2012gv}.
It is still unclear exactly what effect these merger related phenomena have on the constituent galaxies.
There is observational evidence that cluster mergers trigger star formation \citep[e.g.][]{Miller:2003kx,Owen:2005dx,Ferrari:2005es,Hwang:2009ip}, quench it \citep{Poggianti:2004ca}, or have no immediate effect \citep{Chung:2010ds}.
In addition to enabling the study of baryonic physical phenomena, merging clusters can be used to constrain the dark matter self-interaction cross-section by comparing the location of the dark matter (DM; measured through gravitational lensing) with the location of the collisonal gas and effectively collisionless galaxies \citep[e.g.,][]{Randall:2008hs, Dawson:2012dl}. 
There are seemingly conflicting results where galaxy-dark matter offsets have been observed in some systems (A520 \citep{Jee:2012ij} and the Musket Ball \citep{Dawson:2013vg}) but not in others (the Bullet Cluster, \citealt{Bradac:2006be}, and El Gordo, \citealt{Jee:2014ik}).

In an attempt to resolve some of these apparent discrepancies and properly infer the underlying micro-physics, we have formed the Merging Cluster Collaboration\footnote{\url{http://www.mergingclustercollaboration.org}} (MC$^2$) which is undertaking a systematic X-ray, broad/narrow band optical, spectroscopic, and radio survey of an ensemble of merging clusters.
In this paper we will present the global galactic properties of \cizans, the first merger of this systematic approach.
Jee et al.~(submitted) present the weak lensing analysis of this system, and Stroe et al.~(submitted) and Sobral et al. (in preparation) present cluster galaxy evolution analyses.

\ciza (a.k.a.~the Sausage) was first discovered by \citet{Kocevski:2007fc} in the second Clusters in the Zone of Avoidance (CIZA) sample, which is a survey of clusters of galaxies behind the Milky Way.  
Its galactic coordinates are (\dms{104}{11}{20.61}, \dms{-05}{06}{15.87}), so it is very near the disk of the Galaxy but away from the bulge.  
This cluster is in a field with high Galactic dust extinction \citep[A$_v$=1.382; ][]{Schlafly:2011iu}, which is likely the reason there have been limited optical studies of the system (with the exception of our ongoing work and \citet{Stroe:2013ew}'s H$\alpha$ study).

\begin{figure*}
\begin{center}
\includegraphics[width=\textwidth]{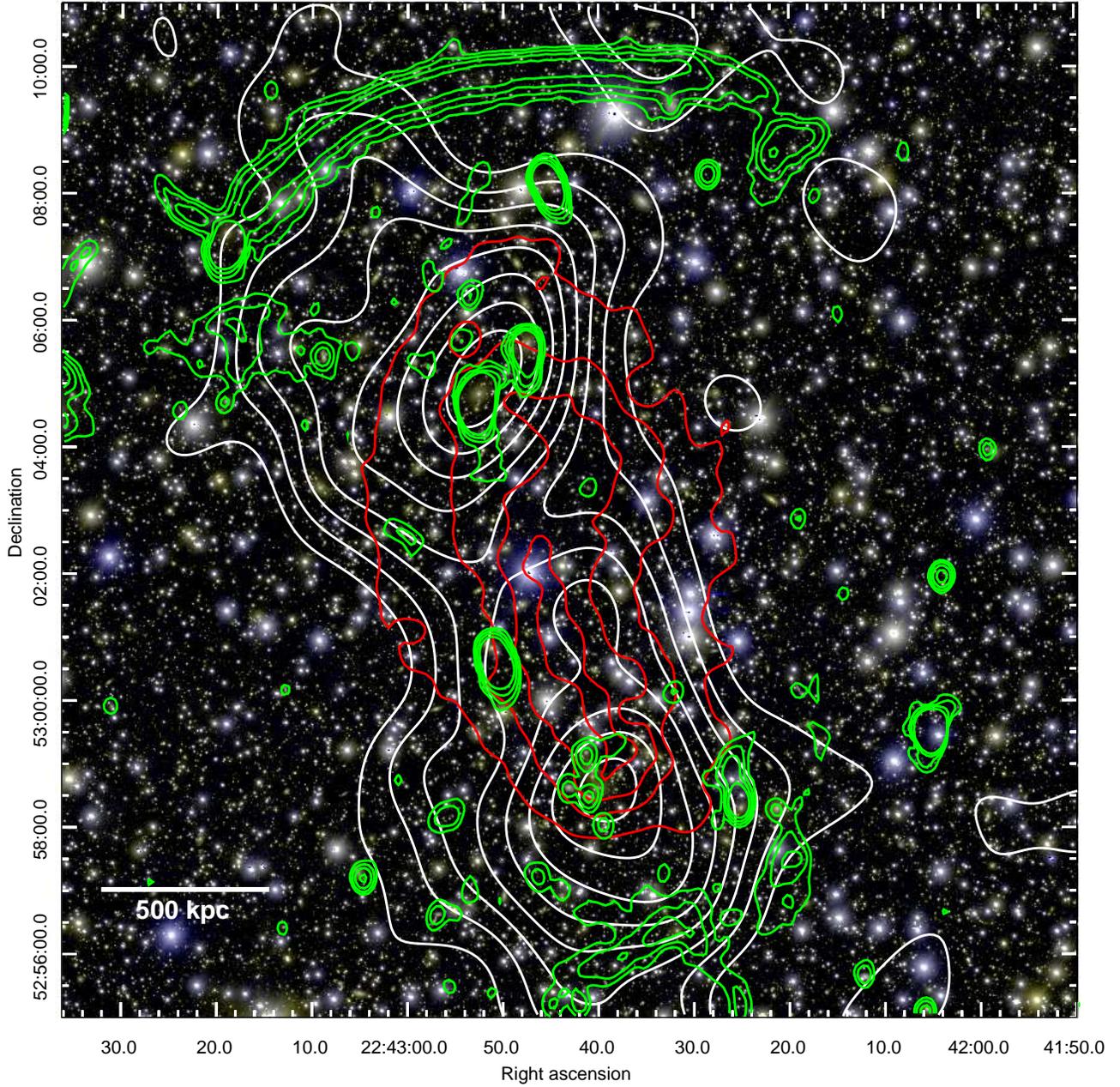}
\caption{Subaru $gi$ color composite image of \cizans.
The red contours are a linear scale mapping of the XMM-Newton X-ray luminosity map.
The green contours are a linear scale mapping of the WSRT radio emission, the radio relics are the extended and diffuse emission near the periphery of the north and south subclusters.
The cluster galaxy number density contours (white) based on our red sequence selection begin at 100\,galaxies\,Mpc$^{-2}$ and increase linearly with increments of 25\,galaxies\,Mpc$^{-2}$ (copied from Figure \ref{fig:DensitywZooms}).
\ciza is an example of a dissociative radio relic merger, with two radio relics at the periphery and the bulk of the cluster gas dissociated between two subclusters.}
\label{fig:SubaruImage}
\end{center}
\end{figure*}

\citet{vanWeeren:2010dn} conducted the first comprehensive radio survey of the system (including Westerbork Synthesis Radio Telescope (WSRT), Giant Meter-wave Radio Telescope (GMRT), and Very Large Array (VLA) observations).
They observed two radio relics at the north and south periphery of the cluster (see green contours in Figure \ref{fig:SubaruImage}).
These radio relics are elongated diffuse radio emission (approximately 10:1 length to width ratios). 
These radio relics are evidence of shock acceleration and spectral aging associated with the outward-moving shock (this was later confirmed with the follow-up study of \citet{Stroe:2013ck}).  
They also observed that the northern relic is strongly polarized at the 50-60\% level, and used this to infer that the merger angle must be within $\sim$30 degrees of the plane of the sky. 
They also used the spectral index to infer a Mach number of $\sim$4.6.

Three detailed X-ray analyses of \ciza have been conducted, one with XMM-Newton \citep{Ogrean:2013gj}, one with Suzaku \citep{Akamatsu:2013tn}, and one with Chandra \citep{Ogrean:2014jo}.
The \citet{Ogrean:2013gj} XMM-Newton analysis shows an extreme N-S elongation of the X-ray gas largely consistent with the merger axis suggested by the radio relics (see red contours of Figure \ref{fig:SubaruImage}). 
The XMM-Newton instrumental background levels prevent them from characterizing the surface brightness profile at the location of the northern radio relic (this is remedied by the \citet{Akamatsu:2013tn} Suzaku observations), however near the southern radio relic they find evidence for a shock with Mach number $\sim$1.2-1.3. 
\citet{Ogrean:2013gj} also note two interesting features of the gas.  
The first feature is a ``wall'' of hot gas east of the cluster center, and while not associated with a radio relic it does extend into the region behind the southern relic.
They note that a simple binary merger is not expected to create such a feature and suggest that it may be indicative of a more complex merger scenario (e.g., a triple merger), or ``a lack of understanding on our part of the complex structures formed during real cluster mergers.'' 
The second feature is a ``smudge'' of enhanced X-ray emission coincident with the eastern ~1/5 of the northern radio relic.  
\citet{Akamatsu:2013tn} found evidence for a temperature jump at the location of the northern radio relic corresponding to a Mach number of $3.15\pm0.52$.
While lower than that estimated by the radio \citep[4.6$\pm$1.3][]{vanWeeren:2010dn}, the Mach number estimates are consistent within the 68\% confidence intervals. 
\citet{Akamatsu:2013tn} did not see a jump in the surface brightness profile, but they claim that this is due to the large Suzaku PSF ($\sim$380\,kpc) being much larger than the width of the relic ($\sim$55\,kpc).
\citet{Ogrean:2014jo} found evidence for two inner density discontinuities, trailing the northern and southern radio relics by $\sim$0.5\,Mpc.
They argue that these discontinuities are not likely cold fronts given that their large distance from the cluster center ($\approx1.5$\,Mpc) would make them the most distant cold fronts ever detected.
Additionally, the measured temperature of $\sim$8-9\,keV would make them the hottest of all known cold fronts.
Instead they argue that the inner density discontinuities could be caused by the violent relaxation of dark matter tidal tails that were generated at the far sides of the dark matter halos post-merger.

A number of simulations of the system have been performed.
\citet{vanWeeren:2011cd} conducted a suite of simulations studying potential analogs to the system and argue that \ciza is undergoing a merger in the plane of the sky ($\lesssim10\arcdeg$ from edge-on), with a mass radio of about 2:1, an impact parameter of $\lesssim400$\,kpc, and a core pass that happened about 1\,Gyr ago.
Interestingly they suggest that the southern subcluster should be slightly less massive, given the relative size of the southern relic. 
\citet{Kang:2012jo} conducted diffusive shock acceleration simulations of the Sausage and found that Mach numbers from 2-4.5 were supported depending on the amount of pre-existing cosmic ray electrons. 
However, they question the ability of the merger event to produce such an elongated shock.
\citet{Stroe:2014ue} recently performed a more comprehensive spectral age simulation of the radio relics and find that the radio observations more likely imply a Mach number of 2.9$^{+0.10}_{-0.13}$, bringing the radio inferred Mach number more in line with the X-ray inferred Mach number.

The only thorough optical analysis of \ciza to date was an H$\alpha$ survey conducted by \citet{Stroe:2013ew}. 
They find an order of magnitude boost in the normalization of the H$\alpha$ galaxy luminosity function in the vicinity of the relics, even greater than that of other known mergers at the same redshift.  
One important note is that they made assumptions about the contamination of their cluster H$\alpha$ population.
Stroe et al.~(submitted) have used the redshifts presented in this paper to show that their original assumptions were overly conservative.
Stroe et al.~(submitted) finds an even larger boost than \citet{Stroe:2013ew} based on updated contamination estimates.

In this paper we add to this picture with a broad band optical and spectroscopic analyses of \cizans, which are key components to properly interpreting the merger.
In \S\ref{sec:ObsImg} we discuss our Issac Newton Telescope Wide Field Camera, Subaru SuprimeCam, and Canada-France-Hawaii Telescope (CFHT) Megacam observations.
In \S\ref{sec:ObsSpec} we discuss our Keck DEep Imaging Multi-Object Spectrograph \citep[DEIMOS][]{Faber:2003ev} and William Hershel Telescope (WHT) AF2 spectroscopic observations.
In \S\ref{sec:ClusterMemberSelection} we discuss our spectroscopic and imaging selection of cluster members.
In \S\ref{sec:SubclustID} we discuss our identification of the systems subclusters and in \S\ref{sec:SubclusterProps} we present the galactic properties of those subclusters.
In \S\ref{sec:MergerPicture} we place the current work in context of the X-ray and radio analyses of the system and where necessary revise existing interpretations.
Finally, in \S\ref{sec:Conc} we summarize our results.

We assume a flat $\Lambda$CDM universe with $H_0 = 70\,\mathrm{km}\,\mathrm{s}^{-1}\,\mathrm{Mpc}^{-1}$, $\Omega_M = 0.3$, and $\Omega_\Lambda = 0.7$. At the redshift of the cluster ($z=0.188$), 1 arcmin corresponds to {189\,kpc}.


\section{Observations: Imaging}\label{sec:ObsImg}

We first observed \ciza in the optical using the Wide Field Camera (WFC) on the 2.5\,m Issac Newton Telescope (INT) at the Roque de Los Muchachos Observatory on La Palma.  
We carried out the observations over two nights (2009, October, 06-07), observing the system in the $B, V, R$, and $I$ filters with total exposure times of 12000\,s, 9000\,s, 9000\,s, 9000\,s, respectively. 
The data reduction was carried out with IRAF and the \emph{mscred} package \citep{Valdes:1998uo}. 
Standard bias and flat-field corrections were carried out and the R and I band images were fringe corrected with \emph{rmfringe}. 
As a final step the images were registered to the 2MASS WCS coordinate system and co-added, rejecting pixels above $3.0\sigma_{\mathrm{rms}}$.
The seeing ranged from 1.5$\arcsec$-2$\arcsec$.
This relatively large point spread function (PSF) coupled with the high stellar densities in the low galactic latitude field made it difficult to discriminate between stars and galaxies when we used the imaging for spectroscopic target selection, as discussed in \S\ref{sec:KeckTargetSelection}.

\ciza was observed with CFHT MegaCam during queue scheduling during 2013 July 3-12 in $r$ (P.I. A. Stroe).
The total integration is 24,000\,s, consisting of 40 short (600\,s) exposures. 
The median seeing is $\sim0.74\arcsec$. 
We also observed \ciza with Subaru SuprimeCam on 2013 July 13 in $g$ and $i$ (P.I. D. Wittman).
The total integration time 700\,s in $g$, consisting of four 180\,s exposures, and a total integration time of 3060\,s in $i$, consisting of eight 360\,s and one 180\,s exposures.
We rotated the field between each exposure (30\,degrees for $g$ and 15\,degrees for $i$) in order to distribute the bleeding trails and diffraction spikes from bright stars azimuthally and later removed them by median-stacking different visits. 
This scheme enables us to maximize the number of detected galaxies.
The median seeing for $g$ and $i$ images are $0.72\arcsec$ and $0.65\arcsec$, respectively.
The details of the CFHT and Subaru data reduction and photometric dust extinction correction are presented in Jee et al.~(submitted).


\section{Observations: Spectroscopic}\label{sec:ObsSpec}

	\subsection{Keck DEIMOS Observations}

\begin{figure*}
\begin{center}
\includegraphics[width=\textwidth]{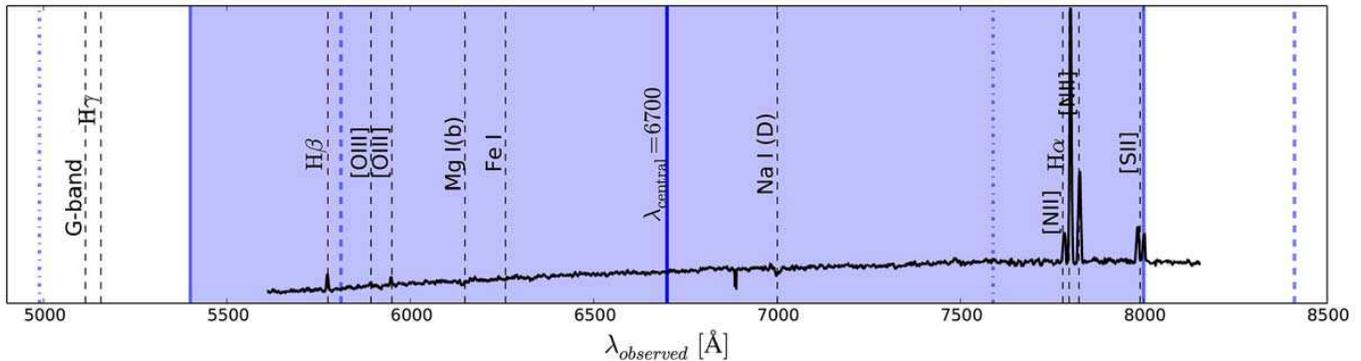}
\caption{
Spectral coverage of the Keck DEIMOS observations (shaded blue region), along with the redshifted location of common cluster emission and absorption features (black dashed lines).  The blue dot-dash pair and the blue dashed pair of lines show the variable range depending on where the slit was located along the width of the slitmask.
The solid black line shows an example galaxy spectrum from our DEIMOS survey.
}
\label{fig:SpecCoverage}
\end{center}
\end{figure*}

We conducted a spectroscopic survey of \ciza with the DEIMOS instrument on the Keck II 10\,m telescope over two observing runs on 2013, July 14 and 2013 September 05. 
Both observing runs were taken with 1\arcsec\ wide slits with the 1200\,line\,mm$^{-1}$ grating, tilted to a central wavelength of 6700\,\AA, resulting in a pixel scale of $0.33$\,\AA\,pixel$^{-1}$, a resolution of $\sim1$\,\AA\ (50\,km\,s$^{-1}$), and typical wavelength coverage of 5400\,\AA\ to 8000\,\AA, shown in Figure \ref{fig:SpecCoverage}.
The actual wavelength coverage may be shifted by $\sim\pm410$\AA\ depending where the slit is located along the width of the slitmask.
For most cluster members this enabled us to observe H$\beta$, [\ion{O}{3}] 4960 \& 5008, \ion{Mg}{1} (b), \ion{Fe}{1},  \ion{Na}{1} (D), [\ion{O}{1}], H$\alpha$, and the [\ion{N}{2}] doublet (Figure \ref{fig:SpecCoverage}).
This spectral setup enables us to also study the star formation properties of the cluster galaxies; see related work by Sobral et al.~(in preparation).
The position angle (PA) of each slit was chosen to lie between $\pm5\degr$ to 30$\degr$ of the slitmask PA to achieve optimal sky subtraction\footnote{\url{http://astro.berkeley.edu/~cooper/deep/spec2d/slitmask.html}} during reduction with the DEEP2 version of the spec2d package \citep{Newman:2012ta}.
Within this range the slit PA was chosen to minimize the effects of chromatic dispersion by the atmosphere by aligning  the slit, as much as possible, with the axis connecting the horizon, object and zenith \citep[see e.g.][]{Filippenko:1982br}.
We observed a total of four slit masks with approximately 120 slits per mask.
For each mask we took three 900\,s exposures.

Since the central wavelength of 6700\,\AA\ is bluer than typical DEIMOS setups we found it necessary to modify the default DEIMOS arc lamp calibration procedure.
We began by turning on the Hg, Ne, Cd, Kr, Ar, and Zn lamps on at the same time, after 1\,s we turned off the Hg and Ne lamps, after 7\,s we turned off the Cd lamp, after 8\,s we turned off the Kr lamp, and stopped exposing after 16\,s.
We found this sequence necessary to prevent the brighter emission lines on the red side from saturating while exposing long enough to get lines of sufficient signal on the blue side.

		\subsubsection{Keck DEIMOS: Target Selection}\label{sec:KeckTargetSelection}

Our primary objective for the spectroscopic survey was to maximize the number of cluster member spectroscopic redshifts.
Since the SuprimeCam imaging was unavailable at the time of our spectroscopic survey planning, we used the WFC imaging to determine the approximate red sequence of the cluster and create a galaxy number density map.
The DEIMOS $5\arcmin\times 16.7\arcmin$ field-of-view (FOV) is well suited to survey the elongated \ciza system, $\sim 7\arcmin \times 15\arcmin$, and we aligned the long axis of our slitmasks with the long axis of the system. 
Cluster member target selection was challenging due to the low galactic latitude ($b =-5\deg$) with a stellar surface density approximately 2.7 times the galaxy surface density, as well as variable extinction ($\Delta E(B-V) \sim$0.4 to 0.6\,magnitudes) across the field \citep{Schlafly:2011iu,Stroe:2013ew}.
The difficulty of star-galaxy separation is also compounded by the 1.5$\arcsec$-2$\arcsec$ seeing of the INT/WFC imaging which results in many stars being blended (especially binary pairs) which results in many blended pairs of stars passing morphological cuts designed to eliminate point sources.
We find that the majority of the stars in the field are bluer than the cluster galaxy population, thus we did not target any object with $R-I<0.9$.
We found it difficult to clearly define the cluster red sequence due to variable extinction across the field plus the red star contamination.
Thus rather than exclude galaxies redder than the brightest cluster galaxy (BCG; $R-I=1.2$) we linearly down weighted the probability of selecting galaxies redder than the BCG as a function of their $R-I$ color.
In addition to these weights, we weighted each galaxy's probability of being targeted by $10^{-(R-22)}$, thus preferentially selecting brighter galaxies likely to have higher signal-to-noise ratios (SNR).
We then divided our potential targets into a bright sample (Sample 1; $R<$22.5) and a faint sample (Sample 2; 22.5$<R<$23.5).
We first filled our mask with as many Sample 1 targets as possible, then filled in the remainder of the mask with Sample 2 targets.

We used the DSIMULATOR package\footnote{\url{http://www.ucolick.org/~phillips/deimos_ref/masks.html}} to design each slitmask.
DSIMULATOR automatically selects targets by maximizing the sum total weights of target candidates, by first selecting as many objects from Sample 1 as possible then filling in the remaining area of the slitmask with target candidates from Sample 2.
We manually edited the automated target selection to increase the number of selected targets, e.g. by selecting another target between targets selected automatically by DSIMULATOR if it resulted in a small loss of sky coverage to their slits.

While we preferentially targeted likely red sequence cluster members it was not always possible to fill the entire mask with these galaxies, in which case we would place a slit on other galaxies in the field.
In our 2013 July 14 observations we serendipitously observed 9 galaxies from the \citet{Stroe:2013ew} H{$\alpha$} catalog.
In our 2013 September 05 observations we purposefully targeted targeted 17 galaxies from that catalog.

		\subsubsection{Keck DEIMOS: Data Reduction}\label{sec:KeckDeimosReduction}

The exposures for each mask were combined using the DEEP2 versions of the \emph{spec2d} and \emph{spec1d} packages \citep{Newman:2012ta}.
This package combines the individual exposures of the slit mosaic and performs wavelength calibration, cosmic ray removal and sky subtraction on slit-by-slit basis, generating a processed two-dimensional spectrum for each slit. 
The \emph{spec2d} pipeline also generates a processed one-dimensional spectrum for each slit.
This extraction creates a one-dimensional spectrum of the target, containing the summed flux at each wavelength in an optimized window. 
The \emph{spec1d} pipeline then fits template spectral energy distributions (SED's) to each one-dimensional spectrum and estimates a corresponding redshift.
There are SED templates for various types of stars, galaxies, and active galactic nuclei (AGN).
We then visually inspect the fits using the \emph{zspec} software package \citep{Newman:2012ta}, assign quality rankings to each fit \citep[following a convention closely related to ][]{Newman:2012ta}, and manually fit for redshifts where the automated pipeline failed to identify the correct fit.
An example of one of the reduced spectra is shown in Figure \ref{fig:SpecCoverage} and more are shown in a related \ciza galaxy evolution paper (Sobral et al. in preparation).

	\subsection{WHT AF2 Observations}

We also conducted a separate spectroscopic survey using WHT/AF2.
This survey primarily targeted H$\alpha$ cluster member candidates identified in our narrow band survey of the system \citep[][\& Stroe et al.~submitted]{Stroe:2013ew}.
In total 73 objects were targeted over an area roughly $30\arcmin\times30\arcmin$.
Specific details regarding the target selection and data reduction processes are presented in Sobral et al.~(in preparation). 

	\subsection{Spectroscopic Redshift Catalog}
		
		\subsubsection{DEIMOS Spectroscopic Redshifts}\label{sec:DEIMOSspeczs}

\begin{figure}
\begin{center}
\includegraphics[width=\columnwidth]{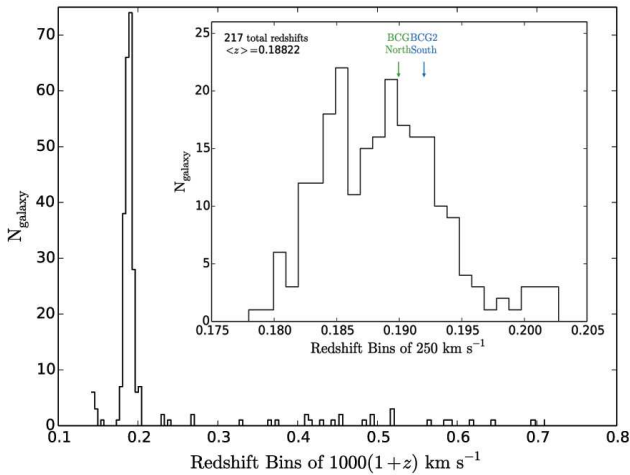}
\caption{
\emph{Main:} Redshift distribution of the Keck DEIMOS high quality (Q $\geq$ 3) galaxy spectroscopic redshifts.
The overdensity near the cluster redshift z=0.188 is clear, with 206 spectroscopic galaxies near the cluster redshift, 15 foreground galaxies, and 34 background galaxies.
\emph{Inset:} A zoom in of the spectroscopic histogram near the cluster redshift.
The north brightest-cluster-galaxy (BCG) redshift is indicated by the green arrow, and the second-brightest southern cluster galaxy is indicated by the blue arrow.
While we targeted the southern BCG we were unable to extract a redshift due to it landing on a chip-gap.
}
\label{fig:hist_deimos_spec}
\end{center}
\end{figure}

We obtained 505 spectra with DEIMOS. Of these, we were able to obtain reliable redshifts for 447 objects (89\%; see Table \ref{tab:Speczs}), leaving 58 spectra which were either too noisy or had ambiguous redshift solutions (e.g., those with a single emission line). 
Figure \ref{fig:hist_deimos_spec} shows the redshift distribution of the 255 (51\%) high quality (Q $\geq$ 3, see \citet{Newman:2012ta} for an explanation on the quality codes) DEIMOS galaxy spectra.
Of the high quality spectra, 206 (41\%) fall within $0.176 \leq z \leq 0.2$, which is  $z_\mathrm{cluster}\pm 3\times\sigma$, where $z_\mathrm{cluster}=0.188$ and $\sigma$ is the approximate velocity dispersion ($1000\,\mathrm{km}\,\mathrm{s}^{-1}$; see \S\ref{sec:zvdisp}).
Of the high quality spectra, 15 (34), or 3\% (7\%), are foreground (background) galaxies and 186 (37\%) of the spectra are stars.
Of these, 72 were serendipitous spectra, meaning that they were not the primary spectroscopic target.
Many of these were paired with primary spectra that were also stars, owing to the fact that often binary stars appeared as a single elliptical in the 1.5$\arcsec$-2$\arcsec$ seeing INT/WFC images.

\begin{deluxetable*}{cccccc}


\tabletypesize{\scriptsize}


\tablecaption{Keck DEIMOS Redshift Catalog of the \ciza Field\label{tab:Speczs}}

\tablehead{\colhead{RA} & \colhead{Dec} & \colhead{z} & \colhead{$\sigma$z} & \colhead{$i$} & \colhead{$\sigma i$} \\
\colhead{(hh:mm:ss.sss)} & \colhead{(dd:mm:ss.ss)} & \colhead{} & \colhead{} & \colhead{(mag)} & \colhead{(mag)} }
\startdata
22:42:43.719 & +52:54:17.317 & 0.187259 & 0.000026 & 18.792 & 0.003\\
22:42:50.008 & +52:54:17.651 & 0.184404 & 0.000057 & 18.949 & 0.003\\
22:42:51.259 & +52:54:22.113 & 0.183752 & 0.000005 & 17.833 & 0.002\\
22:42:36.834 & +52:54:48.770 & 0.454690 & 0.000044 & 19.617 & 0.006\\
22:43:00.240 & +52:54:59.057 & 0.186943 & 0.000023 & 18.650 & 0.002\\
\enddata
\tablecomments{Table 1 is published in its entirety in the electronic
edition of the {\it Astrophysical Journal}.  A portion is shown here
for guidance regarding its form and content.}

\end{deluxetable*}

		\subsubsection{AF2 Spectroscopic Redshifts}
\begin{figure}
\begin{center}
\includegraphics[width=\columnwidth]{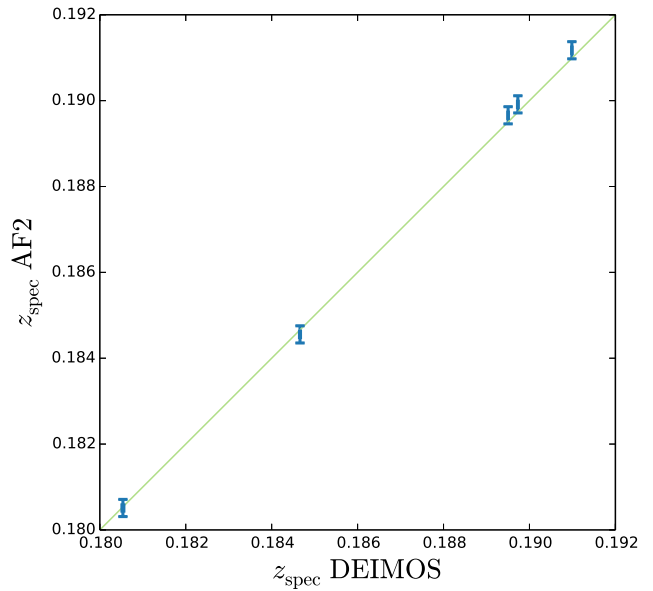}
\caption{
Spectroscopic redshift comparison of the five galaxies in our \ciza survey that have overlapping DEIMOS and AF2 spectra and pass both DEIMOS and AF2 quality cuts (blue error bars).
The green line shows the expected 1:1 ratio.
}
\label{fig:DEIMOSvsAF2}
\end{center}
\end{figure}

We targeted 73 objects with our AF2 spectroscopic survey of cluster H$\alpha$ candidates, 42 of which fall within the 15$\arcmin$ radius of the cluster center being analyzed for this article.
Of those 42 objects 11 (26\%) are stars and 19 (45\%) are high quality galaxy spectra with reliable redshifts.
Five of those galaxies also have high quality DEIMOS spectra (\S\ref{sec:DEIMOSspeczs}).
As can be seen from Figure \ref{fig:DEIMOSvsAF2} we find that the spectroscopic redshifts for the two surveys are consistent within the measurement errors.
In the following analysis we use just the DEIMOS redshift values for these galaxies due to their smaller uncertainties.
Of the 14 unique high quality AF2 redshifts within a 15$\arcmin$ radius of the cluster center, 11 are new cluster members and 3 are higher redshift galaxies.

\section{Galaxy Cluster Member Selection}\label{sec:ClusterMemberSelection}
To determine which galaxies are members of the \ciza cluster we utilize both spectroscopic and red sequence cluster member selection methods. 
The spectroscopic sample has the advantage of a being a more pure sample and the precise redshifts are a necessity for many of the following analyses (see \S\ref{sec:DSTest} \& \S\ref{sec:SubclusterProps}).
While the red sequence sample is not as pure it is more complete and is not subject to the under sampling bias that affects the spectroscopic sample (\S\ref{sec:SpeczSelection}) thus it is advantageous for some analyses (see \S\ref{sec:ProjDensities}).
In this section we define each sample and quantify the sample completeness and purity.

In what follows we limit our consideration to galaxies within 15$\arcmin$ of the center of the Subaru field, RA = \hms{22}{42}{43.762}, Dec = \dms{53}{02}{06.3}.
Subaru SuprimeCam is strongly vignetted beyond this radius, with the corner pixels receiving approximately half the light as the center \citep{vonderLinden:2012vq}.
Since the cluster fits well within this radius there would be little gained by including galaxies outside this radius.

	\subsection{Spectroscopic Redshift Selection}\label{sec:SpeczSelection}
	
All spectroscopic galaxies within the range $0.176 \leq z \leq 0.2$ are considered to be cluster members.
This range is defined by  $z_\mathrm{cluster}\pm 3\times\sigma$, where $z_\mathrm{cluster}=0.188$ and $\sigma$ is the approximate velocity dispersion of each subcluster ($1000\,\mathrm{km}\,\mathrm{s}^{-1}$; see \S\ref{sec:zvdisp}).
This is not exactly a $3\sigma$ selection cut, since the velocity dispersions of the northern and southern subclusters are $1160\pm 95\,\mathrm{km}\,\mathrm{s}^{-1}$ and $1080\pm90\,\mathrm{km}\,\mathrm{s}^{-1}$, respectively, and they have a line-of-sight relative velocity difference of $-69\pm190\,\mathrm{km}\,\mathrm{s}^{-1}$ (see \S\ref{sec:zvdisp}).
This selection results in 206 Keck DEIMOS spectroscopic redshifts and 11 unique WHT AF2 spectroscopic redshifts, for a total of 217 spectroscopic cluster member redshifts. 

\begin{figure}
\begin{center}
\includegraphics[width=\columnwidth]{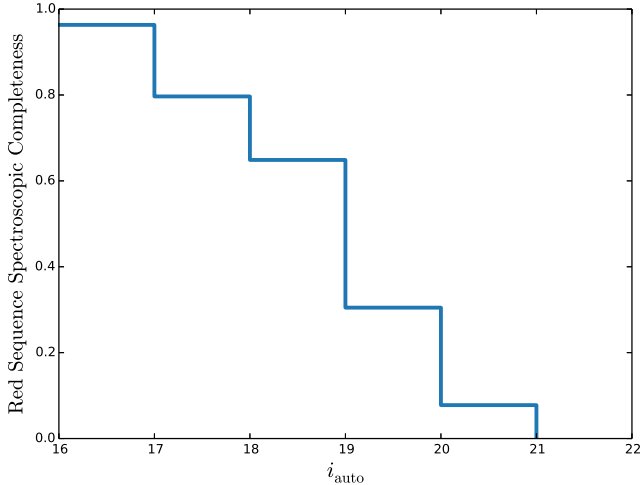}
\caption{
Estimate of the Keck DEIMOS spectroscopic completeness of the cluster red sequence galaxies as a function of extinction corrected Subaru $i$-band magnitude in the Keck DEIMOS survey footprint area.
}
\label{fig:DeimosSpecCompleteness}
\end{center}
\end{figure}

Since our Keck DEIMOS spectroscopic survey targeted primarily cluster red sequence galaxies (see \S\ref{sec:KeckTargetSelection}) it is an incomplete survey of the cluster blue cloud galaxies. 
While the WHT AF2 survey adds a number of blue-cloud galaxies there are only 11 unique additional spectra in the 15$\arcmin$ radius surrounding the cluster.
Also, since the blue cloud region of color-magnitude space has a large amount of stellar contamination it is difficult to estimate our completeness of this population of cluster galaxies.
However, we are able to use the cluster red sequence to estimate our spectroscopic completeness in this region of color-magnitude space.
After correcting for the purity of our red sequence imaging selection (\S\ref{sec:RedSeqProps}) and Keck DEIMOS survey area, we estimate the spectroscopic completeness for red sequence cluster galaxies as a function of $i$-band magnitude, Figure \ref{fig:DeimosSpecCompleteness}.

While our Keck DEIMOS spectroscopic survey is a $\gtrsim$70\% complete sample of cluster red sequence galaxies with $i<$19 (mass $\gtrsim10^{10}$\,M$_\sun$), it is important to note the undersampling bias that affects the densest parts of the subclusters.
Since Keck DEIMOS utilizes slitmasks and the reduction software (\S\ref{sec:KeckDeimosReduction}) is not designed for slits that overlap in the dispersion direction, we undersample the dense cores of each subcluster (see e.g., the insets of Figure \ref{fig:DensitywZooms}) relative to the less-dense periphery of each subcluster.
This bias will affect the southern subcluster more than the northern subcluster, due to its higher galaxy concentration.
	
	\subsection{Red Sequence Selection}\label{sec:RedSeqSelection}

Despite our spectroscopic survey being a $\gtrsim$70\% complete sample of cluster red sequence galaxies with $i<$19, we are able to obtain a more complete and less biased survey of cluster members through color-magnitude selection.
In this subsection we first discuss our star/galaxy separation schema and then discuss our red sequence cluster membership selection schema, as well as the purity of this sample.

		\subsubsection{Star-galaxy Separation}

\begin{figure}
\begin{center}
\includegraphics[width=\columnwidth]{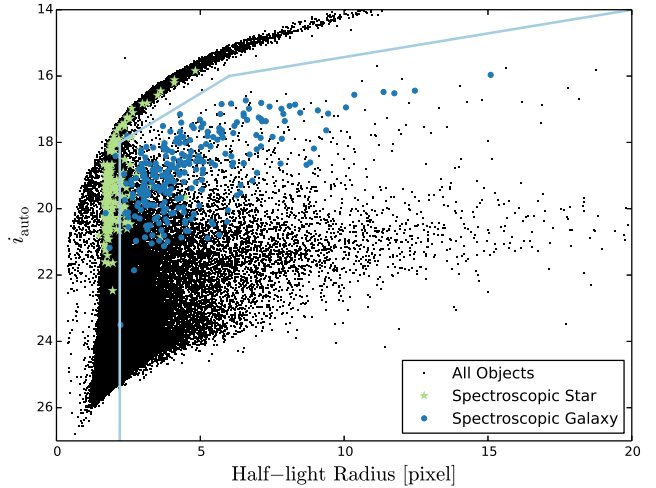}
\caption{
A size magnitude diagram based on Subaru extinction corrected $i$-band magnitude and half-light radius.
Spectroscopically confirmed stars (green stars) and galaxies (blue circles) are overlaid.
The stellar track is visible to the left and above the light blue lines which designate our defined star/galaxy separation border.
For $i>$18 stars are defined to have half-light radii $<$ 2.2 pixels, at $i$=18  the slope changes to -0.53, and at $i$=16 changes to -0.14 in order to track the changing stellar sequence slope due to saturation.
A half-light-radius of 2.2 pixels is 0.44\arcsec for SuprimeCam.
Several spectroscopic stars have half-light radii greater than 2.2 pixels due to blending with neighboring objects.
}
\label{fig:SizeMagDiag}
\end{center}
\end{figure}

The excellent 0.65$\arcsec$ seeing of the Subaru $i$-band imaging facilitates star-galaxy separation via size (or half-light radius) cuts.
We couple this with each objects' magnitude to perform a size-magnitude cut to distinguish between stars and galaxies (see Figure \ref{fig:SizeMagDiag}).
In Figure \ref{fig:SizeMagDiag} we overlay spectroscopically confirmed stars and galaxies as well as our defined border between the star-galaxy phase space.
For $i>$18 stars are defined to have half-light radii $<$ 2.2 pixels (0.44$\arcsec$), at $i$=18  the slope changes to $-0.53$, and at $i$=16 changes to $-0.14$ in order to track the changing stellar sequence slope due to saturation.
Our star-galaxy separation schema errs more towards galaxy completeness than purity, since blending results in a large number of stars with measured half-light radii $>$ 2.2 pixels.

We also investigated whether a color-magnitude cut would increase our star-galaxy discriminating power.
All reasonable color-magnitude cuts resulted in a sample of stars that were already subsumed by the size-magnitude selected sample.

		\subsubsection{Red Sequence Properties}\label{sec:RedSeqProps}

We find that after star-galaxy separation and dust extinction corrections that there is a well defined and relatively tight cluster red sequence (see Figure \ref{fig:ColorMagDiag}; see Appendix \ref{sec:DustComparison} for discussion of the  dust extinction corrections).
We further accentuate this by plotting spectroscopically confirmed cluster members in this color-magnitude space (green points in Figure \ref{fig:ColorMagDiag}).
Across the 15$\arcmin$ radius field there are 2605 presumed galaxies that fall within our defined red sequence region.
We estimate the purity of a cluster red sequence selected sample by studying the population of spectroscopic stars and galaxies within our red sequence selection region.
Our red sequence selection region extends to $i$=22 and our spectroscopic sample completeness falls to $<$10\% beyond $i>$20 (see Figure \ref{fig:DeimosSpecCompleteness}) so our purity calculations should be considered rough estimates.
Within the red sequence selection region there are 234 spectroscopic objects with secure redshifts: 179 (77\%) are cluster galaxy members, 4 (2\%) are foreground galaxies, 14 (6\%) are background galaxies, and 38 (16\%) are stars.
Thus our red sequence selection sample is reasonably pure.
We also find no evidence for clustering of the contaminants and thus expect no resulting propagation of bias in our subcluster location estimates.
We do not attempt to estimate the completeness of this cluster red sequence membership selection schema since our spectroscopic survey was not a magnitude limited survey, instead targeting primarily red sequence galaxies.

\begin{figure}
\begin{center}
\includegraphics[width=\columnwidth]{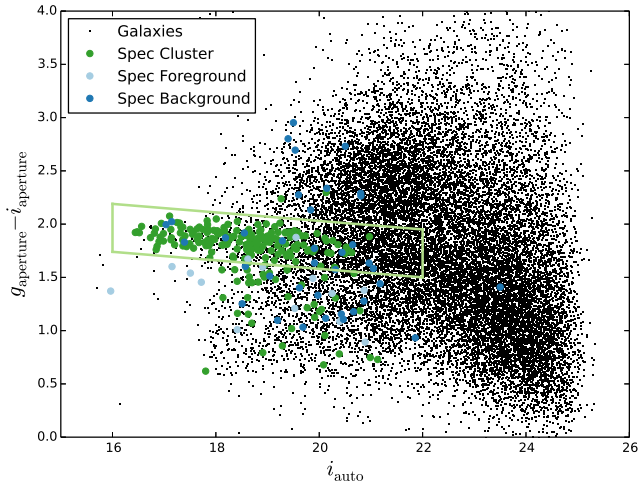}
\caption{
Color-magnitude diagram of galaxies within a 15$\arcmin$ radius of the system center, based on dust corrected Subaru $g$ and $i$ magnitudes.
Spectroscopic cluster (green), foreground (light blue), and background (dark blue) galaxies are overlaid.
Our red sequence selection region is outlined in light green.
}
\label{fig:ColorMagDiag}
\end{center}
\end{figure}


\section{Subcluster Identification}\label{sec:SubclustID}

We employ two separate methods of subcluster identification based on galaxy cluster membership discussed in \S\ref{sec:ClusterMemberSelection}.
The first is the \citet{Dressler:1988co} redshift analysis and the second is a projected galaxy number/luminosity over-density analysis.
These two methods are complementary.
The DS-test has been shown to be one of the best at identifying substructure in cluster \citep{Pinkney:1996ks, Einasto:2012ea}, however it has a notable weakness when attempting to identify substructures with very similar redshifts and velocity dispersions (e.g., subclusters of similar mass that are merging near the plane of the sky).
While the projected galaxy number/luminosity over-density method largely ignores redshift information (except in broad cluster membership selection) it is best at identifying substructure with large relative projected separations \citep[e.g., subclusters that are merging near the plane of the sky][]{Pinkney:1996ks}.

	\subsection{Dressler--Shectman Test}\label{sec:DSTest}

In an attempt to identify the main subclusters and minor substructures in the system, we perform a DS-test \citep{Dressler:1988co} analysis where we calculate the DS-$\delta$ value of each spectroscopic cluster member (see \S\ref{sec:SpeczSelection}).
For each galaxy the DS-$\delta$ parameter is calculated as,
\begin{equation}
\delta^2 = \frac{N_\mathrm{local}}{\sigma^2} \left[\left(\bar{v}_\mathrm{local}-\bar{v}\right)^2+\left(\sigma_\mathrm{local}-\sigma\right)^2\right],
\end{equation}
where $N_\mathrm{local}$ is the number of nearest neighbors (including the galaxy itself) to include when calculating $\bar{v}_\mathrm{local}$, the average line-of-sight velocity, and $\sigma_\mathrm{local}$, the local velocity dispersion.
We let $N_\mathrm{local} = \lceil\sqrt{N_\mathrm{total}}\rceil$, where $N_\mathrm{total}$ is the total number of spectroscopic cluster members, following the best-practice identified by \citet{Pinkney:1996ks}.
Cluster substructures will have larger $\delta$ values.

\begin{figure}
\begin{center}
\includegraphics[width=\columnwidth]{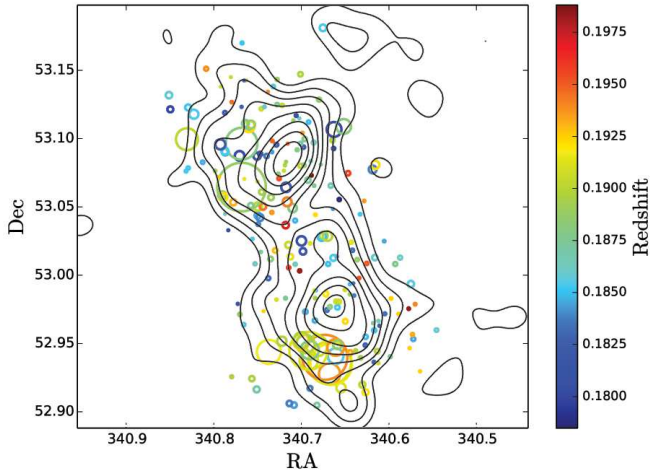}
\caption{
Projected locations of Keck DEIMOS spectroscopic cluster members color coded according to their redshift, with red sequence sample number density contours (see Figure \ref{fig:DensitywZooms} for detailed description).
The diameter of each circle is proportional to 10$^{\delta}$, where $\delta$ is the DS-$\delta$ value for each galaxy: the larger the circle the more likely that galaxy belongs to a substructure with disparate velocity and/or velocity dispersion from that of the bulk system properties.
There are 15 spectroscopic galaxies in the south that show signs of constituting a substructure (i.e., clustering of large circles) with $z\sim0.191$.
Note that the DS-test is not expected to identify the larger north and south subclusters because they have nearly identical velocity distributions.
}
\label{fig:dstest}
\end{center}
\end{figure}

\begin{figure}
\begin{center}
\includegraphics[width=\columnwidth]{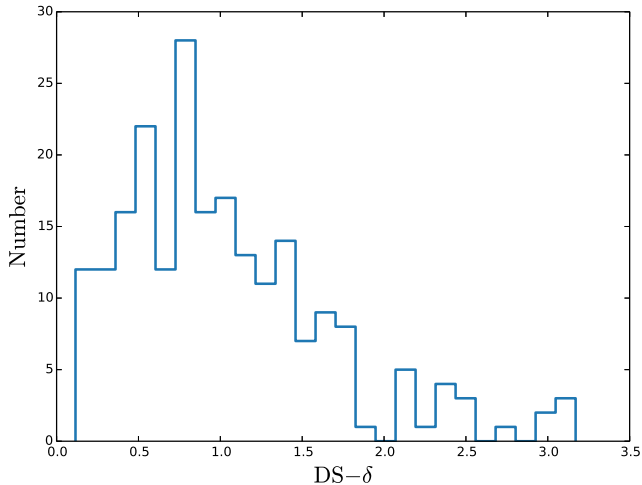}
\caption{
DS-$\delta$ distribution for Keck DEIMOS spectroscopic cluster members.
All but three of the galaxies with $\delta>2.0$ are compactly clustered in the in the south.
We define the galaxies to make up the \emph{interloper}.
}
\label{fig:DSdelta_hist}
\end{center}
\end{figure}

In Figure \ref{fig:dstest} we plot the projected location of each spectroscopic cluster member and represent it as a circle with diameter proportional to 10$^{\delta}$.
We find that there is a concentration of galaxies with large $\delta$ values in the south, indicative of cluster substructure with a mean line-of-sight velocity and/or velocity dispersion different from that of the system average.
Looking at the distribution of $\delta$ values (Figure \ref{fig:DSdelta_hist}) we find that there is a natural break near $\delta$=2.0.
Fourteen of the galaxies with $\delta>2.0$ are compactly clustered in the south.
These make up a small fraction of the total number of spectroscopic galaxies (206) and are considerably fainter than the more massive galaxies that define the southern subcluster peak.
This leads us to define the galaxies as members of a substructure we call \emph{Interloper}.

We investigate the significance of the interloper substructure by comparing  the cumulative deviation,
\begin{equation}
\Delta = \sum_i^{N_\mathrm{total}} \delta_i,
\end{equation} 
of the observed system with that of 10,000 realizations where we maintain the projected galaxy locations but shuffle the redshifts.
When we do this for all of the cluster redshifts we find $\Delta=221.6$ which is only a 0.55$\sigma$ deviation from the mean of the distribution defined by the 10,000 resamplings.
If we instead investigate the significance of the interloper by considering only  redshifts within 625\,kpc of the peak of the southern subcluster (\S\ref{sec:SubclustLoc}) we find$\Delta=93.0$, which is a 1.8$\sigma$ deviation.
Since there is only marginal evidence for the interloper being a distinct substructure, in \S\ref{sec:SubclusterProps}, we consider both the cases where the interloper is a distinct substructure and where the interloper galaxies are just members of the southern subcluster.

With the exception of the interloper galaxies the redshift distributions in the north and south regions of the system are similar, see Figure \ref{fig:dstest}.
The DS-test, designed primarily to identify velocity substructure, cannot separate structures with such similar radial velocity distributions. 
Thus the results are not inconsistent with the previous findings of \citet{vanWeeren:2011cd} suggesting that the \ciza system consists of two near equal mass subclusters.

	\subsection{Projected Densities}\label{sec:ProjDensities}

Given the similar redshift distributions (see Figures \ref{fig:dstest} \& \ref{fig:SubclustSpecProps}) and velocity dispersions in the north and south (as we will discuss in \S\ref{sec:SubclusterProps}) and the known failure mechanism of the DS-test we also look for substructures in projected space.
We independently use both the spectroscopic cluster member sample, \S\ref{sec:SpeczSelection}, and the red sequence cluster member sample, \S\ref{sec:RedSeqSelection}.
For each of these samples we study both the projected galaxy number density distribution and the projected luminosity density distribution (essentially the same as the number density except that we weight each galaxy by its observed $i$-band luminosity, assuming it is at the average redshift of the cluster).

We employed kernel density estimation (KDE)\footnote{A more comprehensive discussion of KDE can be found in either \citealt{Feigelson2014} or \citealt{astroMLText}}
to model the structures of the galaxy data. 
We made use of KDE as the number density estimate,
\begin{equation}
\hat{f}({\bf x}, {\bf h}) = \frac{1}{n\prod_{j=1}^p h_j} \sum_{i=1}^n
\left[ \prod_{j=1}^p K\left(\frac{\bf{x}_i - \bf{X}_{ij}}{\bf{h_j}}\right)
\right],
\end{equation} 
where we use $p=2$ as the number of spatial variables, $n$ as the number of galaxies, $\bf{X}_{i} = (X_{i1}, X_{i2})$ as the two spatial values of each galaxy, ${\bf h} = (h_1, h_2)$ the bandwidth for each dimension, and $K$ is the bivariate Gaussian kernel function. 

The most important aspect of performing a KDE is to pick suitable bandwidths  $\bf{h}$ (smoothing length). 
The smaller the bandwidth the greater the variance in the KDE, however the greater the bandwidth the greater the bias.
The consideration for choosing suitable bandwidths is what is known as the bias-variance trade-off. 
We picked our smoothing bandwidth by performing an exhaustive leave-one-out cross-validation \citep{Stone1984} in each dimension, while maximizing the likelihood of fit between our KDE and the data.    
The cross validated score in each dimension ($l$) can be written as: 
\begin{equation}
CV_l(h_i) = \frac{1}{N}\sum_i^n \ln \hat{f}_{-i, kern}(X_i), 
\end{equation}
where we constructed $n$ datasets, with data of the $i$-th galaxy being left out in each dataset and we performed a grid search of suitable $h_i$ values to maximize the score.
When we apply this procedure to the red sequence selected sample we find that the most suitable bandwidths (i.e., those with the maximum $CV$ score) for the RA and Dec are 62$\arcsec$ and  42$\arcsec$ respectively, and when we apply it to the spectroscopic cluster member sample we find 90$\arcsec$ and 67$\arcsec$ for the RA and Dec dimensions, respectively.
To avoid anamorphic distortions in the projected RA-Dec space, we use the smaller of the two bandwidths for each dimension, 42\arcsec for the red sequence sample and 67\arcsec for the spectroscopic cluster member sample.
We choose the smaller of the two bandwidths in each case since this minimizes bias.
While this choice will slightly increase the variance, we have verified that we are still able to maintain subcluster peak density SNR's $>9$ by performing bootstrap error analyses of each map with 1000 resamplings of the respective galaxy populations.
We find general agreement between each of the four resulting density maps.
For the sake of simplicity, in what follows we will consider just the red sequence number density map, however we present the four resulting density maps in Appendix \ref{sec:GalDenComp}.

\begin{figure}
\begin{center}
\includegraphics[width=\columnwidth]{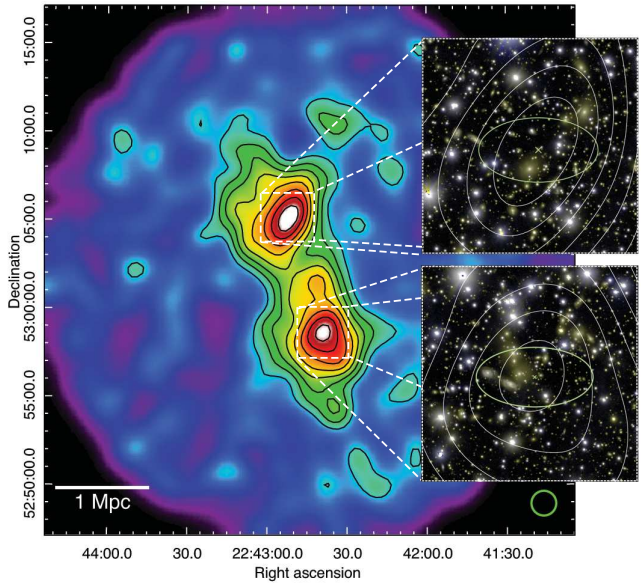}
\caption{
Smoothed galaxy luminosity density map of \ciza based on cluster red sequence selection.
The cluster galaxy number density contours (black) based on our red sequence selection begin at 100\,galaxies\,Mpc$^{-2}$ and increase linearly with icraments of 25\,galaxies\,Mpc$^{-2}$.
Color composites based on the Subaru $g$ and $i$ band observations are shown for the peaks of the north and south subclusters.
The light green ellipses show the 68\% confidence regions for the locations of each subcluster based on 10,000 bootstrap resamplings of the cluster red sequence galaxies.
The dark green circle in the bottom right of the map shows the scale of the KDE bandwidth used to create the map.
}
\label{fig:DensitywZooms}
\end{center}
\end{figure}

From the red sequence number density map presented in Figure \ref{fig:DensitywZooms} it is apparent there are two distinct subclusters (one in the north and one in the south) with similar size and density.
We compare this galaxy density distribution with the X-ray and radio emission of the system in Figure \ref{fig:SubaruImage}.
The two dominant subclusters  that are aligned closely with the merger axis inferred from the radio relics \citep{vanWeeren:2010dn} and elongated X-ray gas distribution \citep{Ogrean:2013gj, Ogrean:2014jo}.
Furthermore, the X-ray gas distribution is largely located between the two galaxy subcluster as expected for a dissociative merger.
We discuss the galaxy distribution in relation to the other cluster emission further in \S\ref{sec:MergerPicture}, and Jee et al. (submitted) discuss the galaxy distribution in relation the the weak lensing mass distribution.

\section{Subcluster Properties}\label{sec:SubclusterProps}

Having defined the northern, southern, and interloper subclusters in \S\ref{sec:SubclustID}, we now present the macroscopic galaxy properties of each subcluster.
Of particular interest are the subcluster locations, redshifts, and velocity dispersions ($\sigma_v$).

	\subsection{Subcluster Locations}\label{sec:SubclustLoc}

Accurate subcluster locations are necessary for the dynamic analysis of the system.
They are also necessary for some constraints on the dark matter self-interactions that rely on accurately constraining the offset between the effectively collisionless galaxies and dark matter.
The dynamic analysis also depends on accurate estimates of each subcluster's redshift, or their relative line-of-sight velocities.
	
To estimate the subcluster locations we use the four KDE projected density maps discuss in \S\ref{sec:ProjDensities}.
We  measure the locations of the north and south subclusters as the peaks of density maps in the north and south regions, respectively.
To estimate the uncertainty distribution on these peak locations we generate 10,000 bootstrap samples from the respective cluster member sample and repeat the same smoothing and peak location process for each, limiting the search region to $\sim$500\,kpc $\times$ 500\,kpc regions surrounding each peak in the original density map. 
We find consistent location estimates in each of the four maps for both the north and south subclusters (see Figure \ref{fig:LocationEB}).
We report here location estimates for the red sequence cluster member sample (\S\ref{sec:RedSeqSelection}) and KDE projected number density map (\S\ref{sec:ProjDensities}), since it is less affected by the spectroscopic undersampling bias (see discussion in \S\ref{sec:SpeczSelection}) and since our bootstrap resampling for the luminosity weighted maps is potentially biased due to resampling the galaxies rather than units of luminosity.
We find that the north subcluster is located at (RA = \hms{22}{42}{50}$^{+50^\mathrm{s}}_{-50^\mathrm{s}}$, Dec = \dms{53}{05}{06}$^{+32\arcsec}_{-23\arcsec}$) and the south subcluster is located at (RA = \hms{22}{42}{39}$^{+50^\mathrm{s}}_{-50^\mathrm{s}}$, Dec = \dms{52}{58}{35}$^{+31\arcsec}_{-18\arcsec}$).
These locations as well as the 68\% confidence regions are shown in Figure \ref{fig:DensitywZooms}.
As can be seen from the north and south zoomed insets of Figure \ref{fig:DensitywZooms} the north and south subcluster peak locations are very near the BCG of each subcluster, 55\,kpc and 85\,kpc, respectively.

Given the north and south subcluster locations we estimate that the projected separation of the two subclusters is $6.9\arcmin^{+0.7}_{-0.5}$, which corresponds to 1.3\,Mpc$^{+0.13}_{-0.10}$.
We estimate the projected separation PDF by selecting 10,000 random samples from the aforementioned north and south subcluster location bootstrap samples and calculate the spherical trigonometric separation of the two in each case.
	
	\subsection{Subcluster Redshifts and Velocity Dispersions}\label{sec:zvdisp}

\begin{figure}
\begin{center}
\includegraphics[width=\columnwidth]{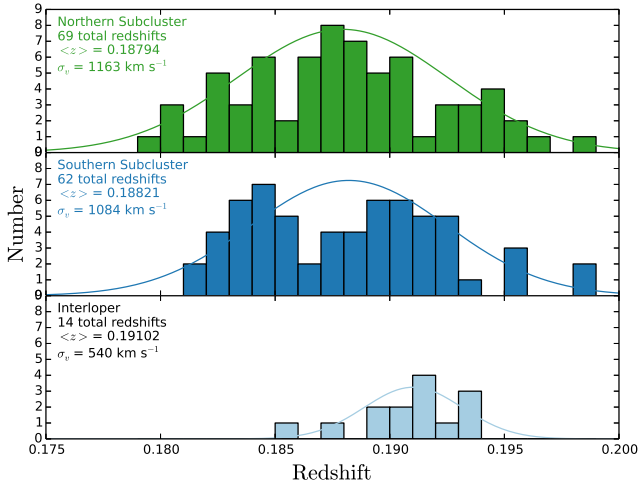}
\caption{
Redshift distributions of the northern subcluster (green), southern subcluster (dark blue), and the potential interloper (light blue).
Redshift locations and velocity dispersions are listed in the upper left of each subpanel. 
The northern and southern subcluster histograms include spectroscopic members within a 625\,kpc radius of the peak location of each subcluster (\S\ref{sec:SubclustLoc}).
Interloper galaxies were excluded from the southern subcluster distribution.
}
\label{fig:SubclustSpecProps}
\end{center}
\end{figure}

To investigate the redshift and velocity dispersions of each subcluster we consider all spectroscopic cluster member galaxies within a 625\,kpc radius of the respective red sequence number density location (see \S\ref{sec:SubclustLoc}).
These apertures were chosen to be as large as possible while maintaining mutual exclusivity of the subcluster membership.
For the southern subcluster we exclude the 14 galaxies identified as interlopers (\S\ref{sec:DSTest}) and estimate their redshift and velocity dispersion separately.
In total we use 69 and 62 redshifts when analyzing the northern and southern subclusters, respectively.
The redshift distributions of each of these selections are shown in Figure \ref{fig:SubclustSpecProps}.
While the southern subcluster redshift distribution appears bimodal there is no sign of corresponding clustering in projected space; as discussed in \S\ref{sec:DSTest}.

We estimate each subcluster's redshift and velocity dispersion using the biweight-statistic and bias-corrected 68\% confidence limit \citep{Beers:1990kg} applied to 100,000 bootstrap samples of each subcluster's spectroscopic redshifts.
We summarize these results in Table \ref{tbl:SubclustProp}.
We find very similar redshifts for the northern and southern subclusters, $0.18794^{+0.00054}_{-0.00054}$ and $0.18821^{+0.00054}_{-0.00052}$, respectively.
These translate to a relative line-of-sight (LOS) velocity difference in the frame of the cluster of  $v_\mathrm{north}-v_\mathrm{south}=-69\pm 190\,\mathrm{km}\,\mathrm{s}^{-1}$. 
This suggests that either they are both nearly in the plan of the sky, have slowed as they near the apocenter of the merger, or a combination of the two.
\citet{vanWeeren:2010dn} argue that the merger is occurring close to the plane of the sky. 
As we will show in a more detailed dynamics analysis (Dawson et al. in preparation) it is likely a combination of the two effects.
Comparing the relative LOS velocities of the southern subcluster and the interloper we find a  larger difference, $v_\mathrm{south}-v_\mathrm{interloper}=-710\pm 200\,\mathrm{km}\,\mathrm{s}^{-1}$, however this is still smaller than the velocity dispersion of the southern subcluster (see Figure \ref{fig:SubclustSpecProps}).

While velocity dispersion mass estimates have been shown to be biased measures in disturbed systems they still provide an independent mass estimate to compare with the less systematic prone weak lensing mass estimates (Jee et al.~submitted).
Furthermore, given the relatively large offset of the subclusters and their relatively small line-of-sight velocity difference we expect that the velocity dispersion bias is not significantly larger than the statistical uncertainty.
We find similar velocity dispersions for the northern and southern subclusters, $1160^{+100}_{-90}\,\mathrm{km}\,\mathrm{s}^{-1}$ and $1080^{+100}_{-70}\,\mathrm{km}\,\mathrm{s}^{-1}$, respectively.
These estimates are consistent with the picture of subclusters of similar richness seen in the galaxy density maps (see e.g., Figure \ref{fig:DensitywZooms}). 
Converting these velocity dispersions into $M_{200}$ mass estimates using the \citet{Evrard:2008jm} scaling relation we estimate masses of $16.1^{+4.6}_{-3.3}\times 10^{14}$\,M$_\sun$ and  $13.0^{+4.0}_{-2.5}\times 10^{14}$\,M$_\sun$ for the northern and southern subclusters, respectively.
These mass estimates are consistent with the Jee et al.~(submitted) weak lensing mass estimates, shown here in Table \ref{tbl:SubclustProp}, although slightly larger.
For the interloper we estimate a velocity dispersion of $540^{+190}_{-110}\,\mathrm{km}\,\mathrm{s}^{-1}$, which translates to $M_{200} = 1.6^{+2.4}_{-0.8}\times 10^{14}\,\mathrm{M}_\sun$.
However as we discussed in \S\ref{sec:DSTest} the evidence for the interloper is not highly significant according to the DS-Test alone.
Even if it is a valid substructure we caution that its velocity dispersion estimate is only based on 14 redshifts and the quoted statistical uncertainties are likely underestimated.

Considering the uncertainty  of the interloper substructure (see \S\ref{sec:DSTest}), we reanalyze the southern subcluster properties including the interloper galaxies to ensure that their exclusion from the analysis of the southern subcluster does not significantly affect our previous findings.
With their inclusion we find $z = 0.18900^{+0.00050}_{-0.00049}$ and $\sigma_v = 1130^{+100}_{-80}\,\mathrm{km}\,\mathrm{s}^{-1}$ for the southern subcluster.
These results are consistent with the results when the interlopers are excluded.	

\begin{deluxetable*}{lcccccc}
\tablewidth{0pt}
\tabletypesize{\small}
\tablecaption{Observed subcluster properties\label{tbl:SubclustProp}}

\tablehead{
\colhead{Subcluster}     & \colhead{RA\tablenotemark{a}} & \colhead{Dec\tablenotemark{a}} & \colhead{Redshift}  &
 \colhead{$\sigma_v$}    &  \colhead{$\sigma_v$ M$_{200}$}&
\colhead{WL M$_{200}$\tablenotemark{b}} \\
\colhead{}         &  \colhead{}  &  \colhead{} &  \colhead{} &
\colhead{(km\,s$^{-1}$)}    &   \colhead{($10^{14}$M$_\sun$)}  & 
\colhead{($10^{14}$M$_\sun$)} 
}
\startdata
North & \hms{22}{42}{50} & \dms{53}{05}{06} & $0.18794^{+0.00054}_{-0.00054}$ & 
 $1160^{+100}_{-90}$ & $16.1^{+4.6}_{-3.3}$ & $11.0^{+3.7}_{-3.2}$  \\
South  & \hms{22}{42}{39} & \dms{52}{58}{35} & $0.18821^{+0.00054}_{-0.00052}$ & 
 $1080^{+100}_{-70}$ &  $13.0^{+4.0}_{-2.5}$ & $9.8^{+3.8}_{-2.5}$\\
Interloper & \hms{22}{42}{43}  & \dms{52}{56}{38} & $0.19102^{+0.00055}_{-0.00065}$  & 
$540^{+190}_{-110}$ & $1.6^{+2.4}_{-0.8}$ & \nodata \\ 
\enddata
\tablenotetext{a}{The uncertainties on these locations are discussed in \S\ref{sec:SubclustLoc}.}
\tablenotetext{b}{As presented in Jee et al.~(submitted) weak lensing analysis of \cizans.}
\end{deluxetable*}


\section{Discussion}

Our findings largely support the general interpretation that \ciza is a major cluster merger being observed sometime after the first pericentric collision.
We have found that \ciza is dominated by two subclusters of similar scale, density, and mass.
Furthermore we are able to accurately locate the subclusters at the leading edges of the elongated X-ray emission, with the bulk of the X-ray emitting gas located between the two subclusters, making this a textbook dissociative merger.

	\subsection{Multi-wavelength Merger Picture}\label{sec:MergerPicture}
	
	
\citet{vanWeeren:2010dn} suggest that the observed radio relic polarization of 50-60\% indicates that the merger angle must be within $\sim$ degrees of the plane of the sky.
Our observed line of sight relative velocity of the subclusters, $v_\mathrm{north}-v_\mathrm{south}=-69\pm 190\,\mathrm{km}\,\mathrm{s}^{-1}$ is consistent with a merger occurring nearly in the plane of the sky, however without further analysis we cannot rule out the possibility that the merger has a larger inclination angle and is just being observed near the merger apocenter where the subclusters have slowed just before or after turn-around.
We will address this in future paper where we present our detailed geometric and dynamic analysis of the system (Dawson et al. in preparation).

\citet{vanWeeren:2011cd} conducted simulations of the system and argue that they suggest a bimodal merger with a 2:1 mass ratio between the north and south subclusters and an impact parameter less than 400\,kpc.
While our velocity dispersion based mass estimates prefer closer to a 1:1 mass ratio, our mass estimates are consistent with a 2:1 mass ratio.
It is difficult for us to place tight constraints on the impact parameter, however we find excellent agreement between the merger axis inferred from the radio relics and elongated gas distribution (see Figure \ref{fig:SubaruImage}), suggesting that the impact parameter is not significantly larger than our uncertainty on the subcluster locations $\lesssim$200\,kpc.

\citet{Ogrean:2013gj} find a ``wall'' of hot gas east of the cluster center.
They note that a simple binary merger is not expected to create such a feature and suggest that it may be indicative a more complex merger scenario (e.g. a triple merger), or ``a lack of understanding on our part of the complex structures formed during real cluster mergers.''
Other than an insignificant substructure we find no evidence for a complex multiple merger scenario and instead our analysis favors a relatively clean bimodal merger.
This then suggest that our understanding of how complex X-ray structures are formed during cluster mergers is incomplete.

As discussed in \S\ref{sec:Intro}, \citet{Ogrean:2014jo} found evidence for two inner density X-ray discontinuities, trailing the northern and southern radio relics by $\sim$0.5\,Mpc.
They argue that these discontinuities are not likely cold fronts given that their large distance from the cluster center ($\approx1.5$\,Mpc) would make them the most distant cold fronts ever detected. However the location of these potential cold fronts should be measured with respect to the two subclusters and not the cluster system center.
In this case the potential cold fronts are offset less than 500\,kpc from the centers of their respective subclusters.
However, this does not help explain the anomalously high temperatures of $\sim$8-9\,keV.
Our observations may also lend support to the possibility that the inner density discontinuities are the result of violent relaxation of the dark matter halos since the galaxies suggest that their are two dark matter halos near the discontinuities.

\citet{Stroe:2013ew} find an order of magnitude boost in the normalization of the galaxy luminosity function in the vicinity of the relics.
Our analysis finds no evidence that this boost in star formation is simply due to coincidental infalling groups or other substructure, suggesting that the enhanced star formation may be the result of cluster members responding to changes in their environment due to the merger.

Our finding of two dominant subclusters, with similar mass based on velocity dispersions is consistent with our weak lensing analysis of the system (Jee et al.~submitted).
In that paper we perform a detailed analysis comparing the relative locations of the galaxies and mass in that paper.



\section{Conclusions}\label{sec:Conc}

We have presented our comprehensive broadband imaging (\S\ref{sec:ObsImg}) and spectroscopic survey (\S\ref{sec:ObsSpec}) of the \ciza cluster's galaxy population, and used this information to provide a new perspective on the cluster's global properties.
We confirm that the cluster is a textbook dissociative major merger, with the bulk of the gas being offset between two galaxy subclusters. 
We also find excellent agreement between the merger axis inferred from the two radio relics, elongated X-ray gas, and bimodal galaxy distribution (see Figure \ref{fig:SubaruImage}).

We find that the system is dominated by two subclusters of comparable richness (\S\ref{sec:SubclustID}) and accurately measure their locations (\S\ref{sec:SubclustLoc}), which imply a projected separation of $6.9\arcmin^{+0.7}_{-0.5}$ (1.3\,Mpc$^{+0.13}_{-0.10}$).
We find that the north and south subclusters have redshifts of $0.18794^{+0.00054}_{-0.00054}$ and $0.18821^{+0.00054}_{-0.00052}$, respectively, corresponding to a relative line-of-sight velocity of $69\pm 190\,\mathrm{km}\,\mathrm{s}^{-1}$.
This is consistent with previous suggestions that the merger is occurring close to the plane of the sky, however without a more detailed dynamics analysis we cannot rule out the possibility that the merger has a larger inclination angle and is just being observed near the merger apocenter.

We also find that north and south subclusters have velocity dispersions of $1160^{+100}_{-90}\,\mathrm{km}\,\mathrm{s}^{-1}$ and $1080^{+100}_{-70}\,\mathrm{km}\,\mathrm{s}^{-1}$, respectively (\S\ref{sec:zvdisp}).
These correspond to masses of $16.1^{+4.6}_{-3.3}\times 10^{14}$\,M$_\sun$ and  $13.0^{+4.0}_{-2.5}\times 10^{14}$\,M$_\sun$, respectively.
While velocity dispersion measurements of merging clusters can be biased we believe the bias in this system to be minor due to the large projected separation and nearly plane-of-sky merger configuration.
In this regard we find the velocity dispersion inferred masses to be consistent with our weak lensing inferred masses (Jee et al.~submitted).
\ciza is a relatively clean dissociative cluster merger, potentially occurring near the plane of the sky, with near 1:1 mass ratio, which makes it an ideal merger for studying merger associated physical phenomena.


\newpage

\section*{Acknowledgments}

We would like to thank the broader membership of the Merging Cluster Collaboration for their continual development of the science motivating this work, which has also been in the acquisition of the data using in this paper.
We would like to thank Anja von der Linden for the initial recommendation to rotate the Subaru SuprimeCam instrument 90\,degrees between exposures to better probe instrument systematic.
Slight modifications to this strategy enabled us to reduce the effects of stellar bleeds and nearly double the number of detected objects.
We also would like to thank Michael Schneider for valuable feedback regarding the presentation of the current work.
 
MJJ, DW, and WD acknowledge support from HST-GO-13343.01-A. 
AS acknowledges financial support from NWO. 
DS acknowledges financial support from the Nether- lands Organisation for Scientific research (NWO) through a Veni fellowship, from FCT through a FCT Investigator Starting Grant and Start-up Grant (IF/01154/2012/CP0189/CT0010) and from FCT grant PEst-OE/FIS/UI2751/2014. 
RW is supported by NASA through the Einstein Postdoctoral grant number PF2- 130104 awarded by the Chandra X-ray Center, which is operated by the Smithsonian Astrophysical Observatory for NASA under contract NAS8-03060.
MB acknowledges support by the research group FOR 1254 funded by the Deutsche Forschungsgemeinschaft. 
MB acknowledges funding from the Deutsche Forschungsgemeinschaft under SFB 676. 
Part of this work performed under the auspices of the U.S. DOE by LLNL under Contract DE-AC52-07NA27344.

This research has made use of NASA's Astrophysics Data System.
The William Herschel Telescope and Isaac Newton Telescope are operated on the island of La Palma by the Isaac Newton Group in the Spanish Observatorio del Roque de los Muchachos of the Instituto de Astrof\'{\i}sica de Canarias. 
Some of the data presented herein were obtained at the W.M. Keck Observatory, which is operated as a scientific partnership among the California Institute of Technology, the University of California and the National Aeronautics and Space Administration. The Observatory was made possible by the generous financial support of the W.M. Keck Foundation.
Funding for the DEEP2/DEIMOS pipelines has been provided by NSF grant AST-0071048.  The DEIMOS spectrograph was funded by grants from CARA (Keck Observatory) and UCO/Lick Observatory, a NSF Facilities and Infrastructure grant (ARI92-14621), the Center for Particle Astrophysics, and by gifts from Sun Microsystems and the Quantum Corporation.

{\it Facilities:} \facility{Keck:II (DEIMOS)}, \facility{Subaru (Suprime-Cam)}, \facility{CFHT (MegaCam)}, \facility{ING:Herschel (AF2)}, \facility{ING:Newton (WFC)}, \facility{XMM}, \facility{WSRT}.

\clearpage

\appendix

\section{Effect of Galactic Extinction on the Color-Magnitude Relation}\label{sec:DustComparison}

As discussed in \citet{Stroe:2013ew} and Jee et al.~(submitted) the Galactic dust extinction in the \ciza field results in high attenuation (A$_v$=1.382), which varies significantly across the field  \citep[$0.35<\mathrm{E}(B-V)<0.52$;][]{Schlafly:2011iu}.
We use extinction values from \citet{Schlafly:2011iu} to recover reddened magnitudes.
Given the spatial resolution of $\sim4\arcmin$, we interpolate between the extinction pierce points using cubic interpolation to predict the dust attenuation at each source position. 
We correct the g, r and i magnitudes by interpolating in wavelength to the effective wavelength of the Subaru and CFHT filters \citep[see][for details]{Stroe:2013ew}.

\begin{figure}
\begin{center}
\includegraphics[width=\columnwidth]{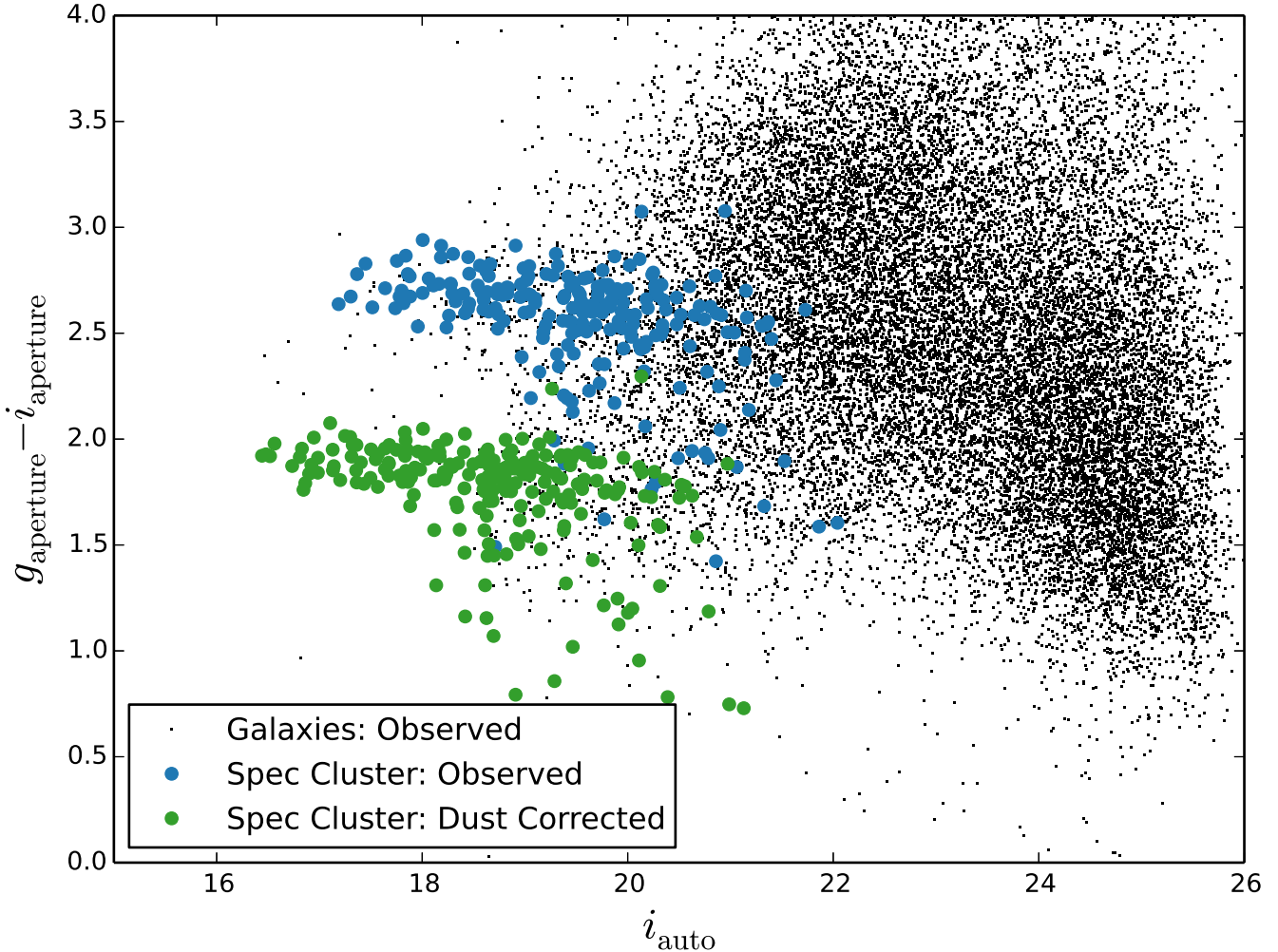}
\caption{
Color-magnitude diagram of galaxies within a 15$\arcmin$ radius of the system center, based on non-dust corrected Subaru $g$ and $i$ magnitudes (\emph{observed}).
Spectroscopic cluster members using non-dust corrected magnitudes are shown in blue.
For comparison the spectroscopic cluster members using dust corrected magnitudes are copied from Figure \ref{fig:ColorMagDiag} and shown in green.
The red sequence is noticeably tighter in color space after applying the dust corrections, which is reflective of the highly varying attenuation across the field.
}
\label{fig:ColorMag_wwoDustCorr}
\end{center}
\end{figure}

In Figure \ref{fig:ColorMag_wwoDustCorr} we plot the color-magnitude diagram using the non-dust corrected magnitudes (\emph{observed}).
For comparison we copy the spectroscopic cluster members using dust corrected magnitudes from Figure \ref{fig:ColorMagDiag} and shown in green.
The red sequence is noticeably tighter in color space after applying the dust corrections, which is reflective of the highly varying attenuation across the field.
Additionally, the translation of the red sequence in the $i_\mathrm{auto}$ direction illustrates the large average attenuation (A$_v\sim0.75$) across the field.

\section{Galaxy Density Comparisons}\label{sec:GalDenComp}

Here we compare the galaxy density maps for the red sequence selection (\S\ref{sec:RedSeqSelection}) and spectroscopic cluster member selection (\S\ref{sec:SpeczSelection}) samples, with and without galaxy luminosity weighting. 
This is an extension of our projected galaxy density discussion in \S\ref{sec:ProjDensities}.

\begin{figure}
\begin{center}
\includegraphics[width=\columnwidth]{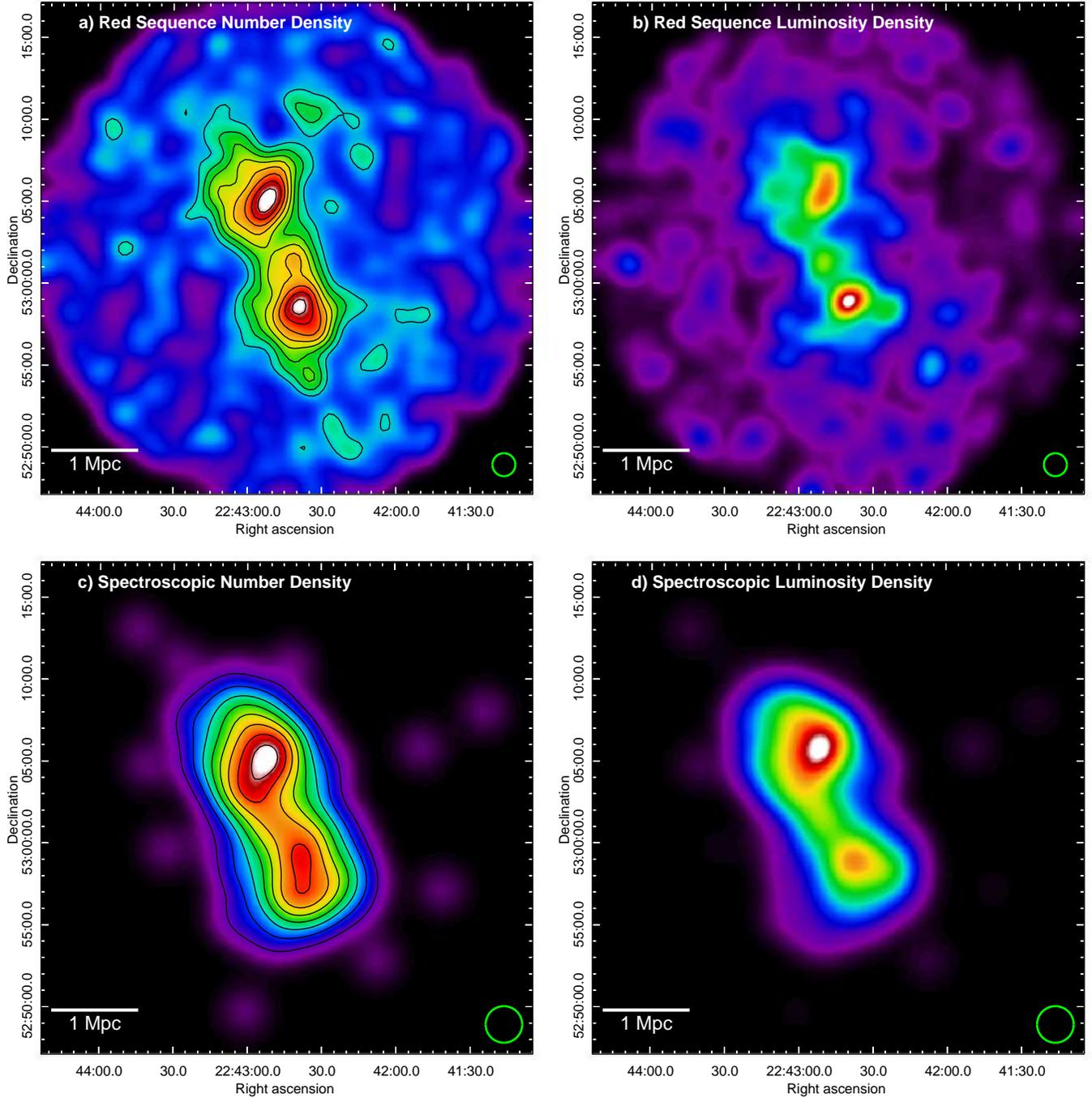}
\caption{
The four galaxy density maps discussed in \S\ref{sec:ProjDensities}.
Each map has a linear scaling with black being less and white being more galaxies\,Mpc$^{-2}$.
The KDE bandwidth (i.e., smoothing scale) is shown by the green circle in the lower right of each map.
The white bar in the lower left gives the physical scale at the cluster redshift $z=0.188$.
\emph{a) }The cluster red sequence sample number density map. 
The cluster galaxy number density contours (black) begin at 100\,galaxies\,Mpc$^{-2}$ and increase linearly with increments of 25\,galaxies\,Mpc$^{-2}$.
The scale has been corrected for contamination (see discussion in \S\ref{sec:RedSeqSelection}). 
\emph{b) } The cluster red sequence sample luminosity density map.
\emph{c) } The cluster spectroscopic sample number density map.
The cluster spectroscopic number density contours (black) begin at 10\,galaxies\,Mpc$^{-2}$ and increase linearly with increments of 10\,galaxies\,Mpc$^{-2}$.
\emph{d) } The cluster spectroscopic luminosity density map.
}
\label{fig:4densitymaps}
\end{center}
\end{figure}

We find that each of these four representations have the same general features (see Figure \ref{fig:4densitymaps}): two dominant subclusters (one in the north and one in the south) that are aligned closely with the merger axis inferred from the radio relics \citep{vanWeeren:2010dn} and elongated X-ray gas distribution \citep{Ogrean:2013gj}, see for example the red sequence sample number density map in Figure \ref{fig:DensitywZooms} (All four maps are shown in Figure \ref{fig:4densitymaps} of the Appendix).
We also find close agreement between the subcluster peak locations for each representation.
There are some notable differences between the maps. 
The number density maps result in slightly less concentrated distributions compared to the luminosity density maps.
The red sequence sample luminosity density map shows that the southern subcluster is more densely concentrated relative to the northern subcluster, see Figure \ref{fig:DensitywZooms}, whereas the spectroscopic sample luminosity density map shows them to be about the same.
This can be explained by spectroscopic undersampling bias (see \S\ref{sec:SpeczSelection} for discussion).
In the zoom of the southern subcluster region in Figure \ref{fig:DensitywZooms} it is apparent that there are a number of very densely packed bright cluster galaxies (compare with the less densely packed northern subcluster peak).
Thus we were able to obtain a more complete spectroscopic survey of the northern subcluster.

\begin{figure}\label{fig:LocationEB}
\plottwo{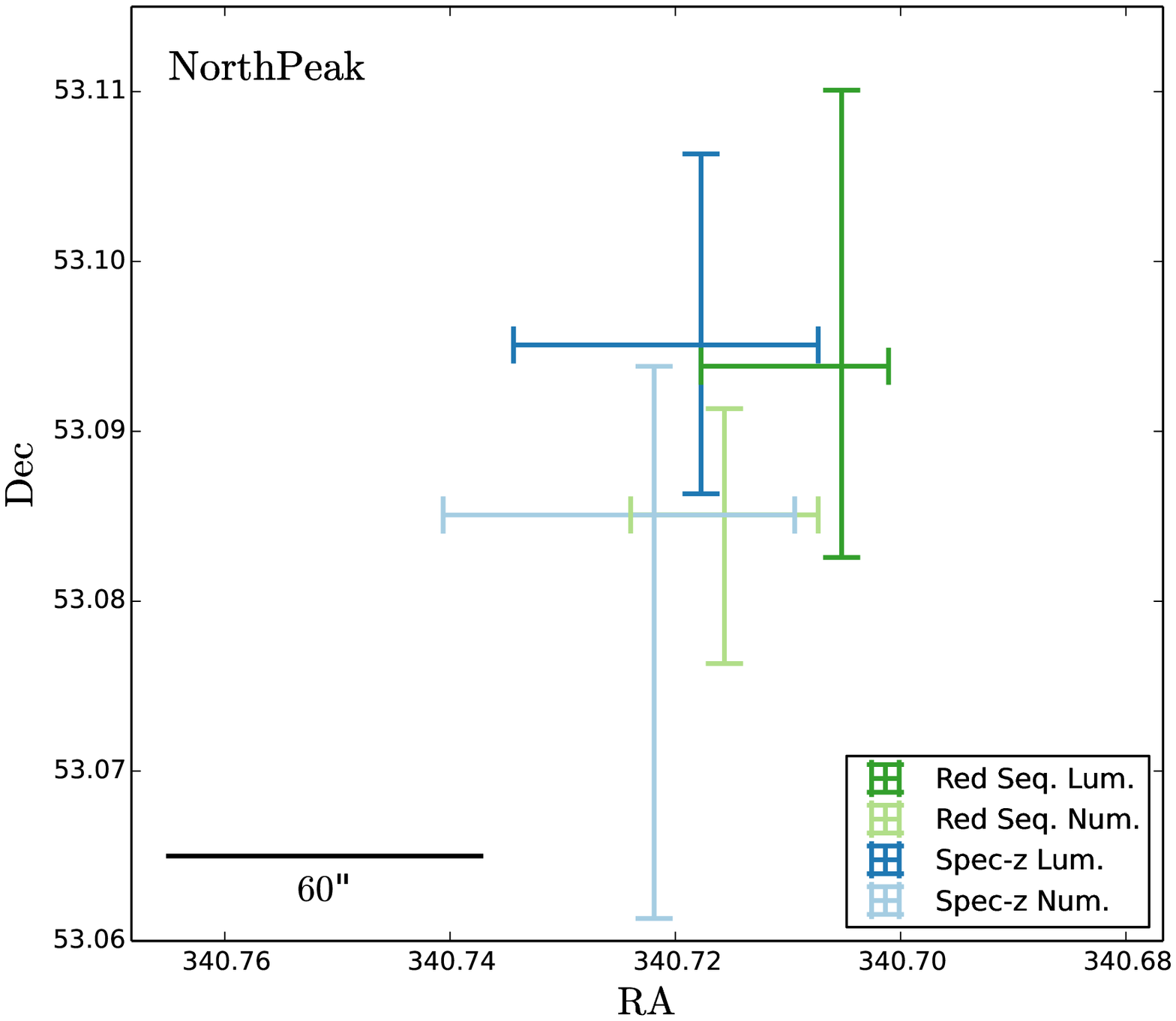}{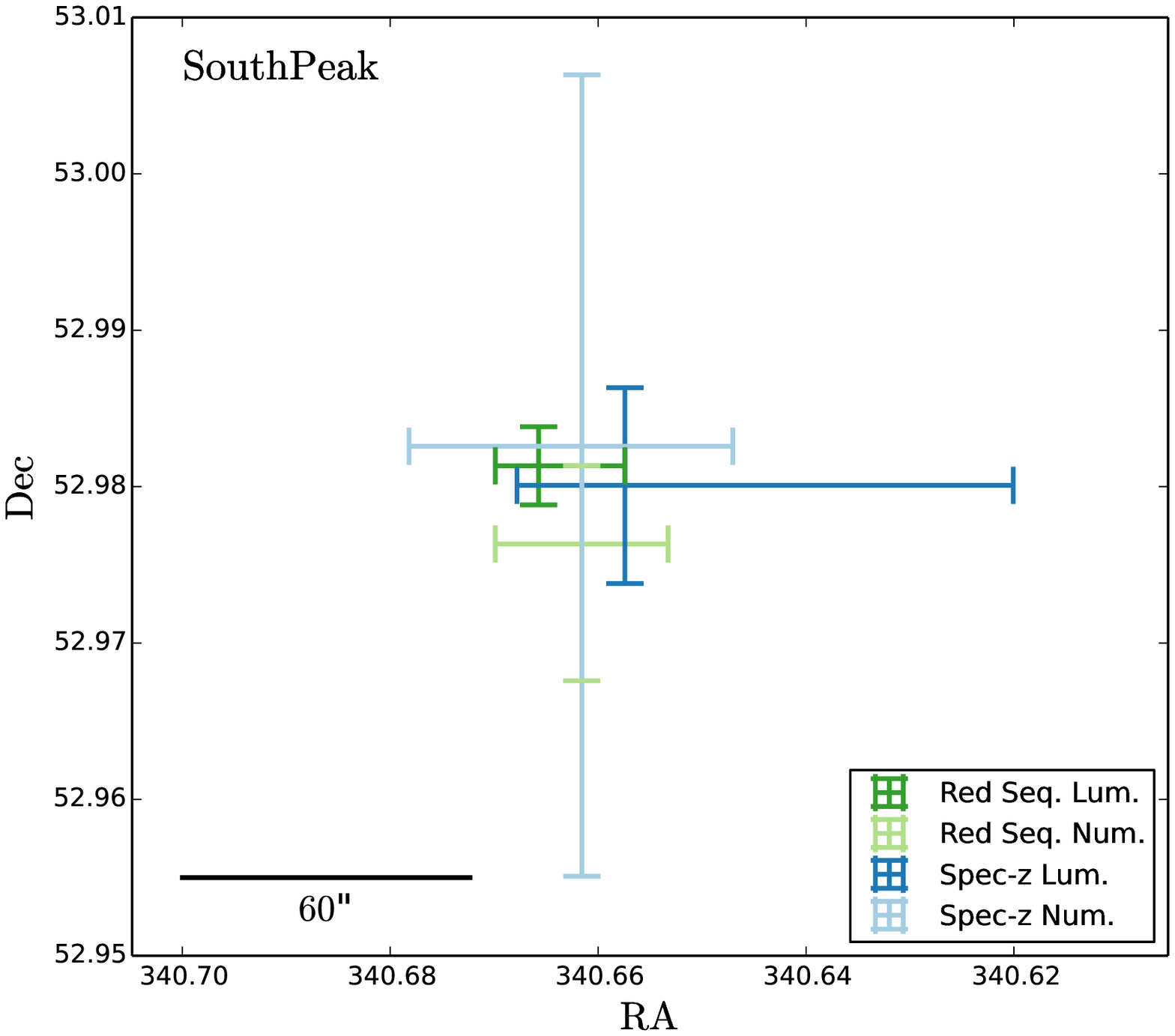}
\caption{A comparison of the peak 65\% confidence intervals for the north (left) and south (right) subclusters, as measured for the red sequence number (light green) \& luminosity (dark green) projected density maps, and spectroscopic cluster sample number (light blue) \& luminosity (dark blue) projected density maps.
The dimensions of each figure are chosen to prevent anamorphic distortions.
We find general agreement between each location measure.
}
\end{figure}

\bibliographystyle{apj}
\bibliography{SausageRedshift}

\end{document}